\documentclass[prl,twocolumn,superscriptaddress,notitlepage,9py]{revtex4-1}
\usepackage[utf8]{inputenc}
\usepackage{natbib}
\usepackage{graphicx}
\usepackage{color}
\usepackage[left=1.5cm, right=1.5cm, top=1.785cm, bottom=2.0cm]{geometry}
\usepackage{floatrow}

\begin{document}

\title{\LARGE Quantification of plasticity via particle dynamics above and below yield in a 2D jammed suspension}

\author{{\Large K.L. Galloway$^{1}$, D.J. Jerolmack$^{1,2}$, P.E. Arratia$^{1}$ \\}
{\large $^{1}$Department of Mechanical Engineering and Applied Mechanics, University of Pennsylvania \\
$^{2}$Department of Earth and Environmental Science, University of Pennsylvania \\ 
}}

\date{\today}

\begin{abstract}
    \normalsize{\textbf{ Failure of amorphous materials is characterized by the emergence of dissipation. The connection between particle dynamics, dissipation, and overall material rheology, however, has still not been elucidated. Here, we take a new approach relating trajectories to yielding, using a custom built interfacial stress rheometer, which allows for  measurement of shear moduli (G',G'') of a dense athermal suspension's microstructure while simultaneously tracking particle trajectories undergoing cyclic shear. We find an increase in total area traced by particle trajectories as the system is stressed well below to well above yield. Trajectories may be placed into three categories: reversibly elastic paths; reversibly plastic paths, associated with smooth limit cycles; and irreversibly plastic paths, in which particles do not return to their original position. We find that above yield, reversibly plastic trajectories are predominantly found near to the shearing surface, whereas reversibly elastic paths are more prominent near the stationary wall. This spatial transition between particles acting as solids to those acting as liquids is characteristic of a 'melting front', which is observed to shift closer to the wall with increasing strain. We introduce a non-dimensional measure of plastic dissipation based on particle trajectories that scales linearly with strain amplitude both above and below yield, and that is unity at the rheological yield point. Surprisingly, this relation collapses for three systems of varying degrees of disorder.}}
\end{abstract}

\maketitle

\section{Introduction}

Much of our natural and built environment is made of amorphous materials, such as foods, foams, and glasses \cite{larson99,chen10}. The properties of disorder may be exploited to create materials with desirable properties, e.g. yield stress or shear thinning fluids. When amorphous solids fail, however, catastrophic fluidization may occur: witness the collapse of solid soil into fast-flowing mudslides \cite{Jerol19, Ander2010}. Therefore, predicting and controlling failure within these materials is of fundamental importance. Bulk rheology of amorphous solids is an emergent property arising from micro-scale grain-grain and fluid-grain interactions \cite{gopal95,hecke10}. There has been major progress in unifying the rheology and yielding of ideal granular materials and suspensions \cite{boyer11,guaz18}. Yet such descriptions are mostly phenomenological; moreover, small variations in particle size, shape, surface properties, or inter-particle forces may cause dramatic changes in bulk properties. Often, constituent particles may be jammed together, preventing all undriven motion either by confinement or by outside forces such as gravity. These factors complicate the unifying description of such materials by creating structure based effects, which commonly cause history dependent responses \cite{hecke10,larson99,Jerol19}.

A key insight has been that as energy is injected via shear into a disordered material, bulk deformations are achieved via contributions from local rearrangements \cite{falk98, Mann11}. These rearrangements within the microstructure are thought of as particles shifting to lower energy configurations, thereby dissipating some of the injected energy \cite{falk11}. The energy not dissipated is recovered as elastic energy. The yield transition is quantified as a shift from mainly elastic to dissipative response with increasing strain \cite{larson99}. For a wide range of disordered materials, a universal strain of $\sim3\%$ has been found to mark the yield transition \cite{cubuk17}. 


A convenient way of repeatedly probing a system's elasticity (storage modulus) and dissipation (loss modulus) is to subject the material to oscillatory stress. This method gives statistically robust measurements over as many cycles as desired. Under oscillatory stress, three types of particle dynamics have been observed. First are those that return to their initial position by the same path they went in (elastic and reversible). Second are those that return via a secondary path (plastic but reversible). Third are particles that do not return at all (plastic and irreversible) \cite{pine05, lund08, slot12, regev13, priez13}. It is thought that reversible particles enter a new minima in the energy landscape, but are returned once strain reverses; i.e. the energy landscape is restored. However, irreversible particles do not return, because of permanent modification to the energy landscape by small perturbations in the positions of neighbors \cite{mobius14,regev15}. It has been observed that reversible, plastic trajectories emerge at the same strain amplitude as the bulk material's rheological yield \cite{keim13,keim14}. Therefore, understanding reversible, plastic trajectories may shed new light on the yield transition in amorphous materials.

Strikingly, reversible, plastic trajectories are similar to a classic limit cycle description of dissipation from non-linear analysis \cite{keim14, Strog94}. Past research has explored this idea, showing via simulations that the area traced is related to energy dissipated \cite{schreck13,regev13}. An intuitive implication is that reversibly plastic trajectories within the same system may share similar properties, reflecting changes to an energy landscape by stress. It may be possible that reversibly plastic particle trajectories are stable from one cycle to the next, corresponding to specific meta-stable states within the energy landscape \cite{kawa16,regev13,regev15,perch14}. More broadly, particle dynamics (reversible vs irreversible, elastic vs plastic) near and above yielding are still not well understood. 

In particular, there are still many outstanding questions regarding how particles transition from the elastic to plastic regime. These include: Are irreversible particle trajectories born out of reversible plastic trajectories (limit cycles)? Or are reversibly plastic particles stable as a function of strain, space, or time? And is there a relationship between the properties of these reversible plastic trajectories and the material's macroscopic rheology? Answering these questions will help elucidate microscopic factors that bring about bulk material yield and inform models. 

In this manuscript, we experimentally investigate the Lagrangian dynamics of particle trajectories in a two-dimensional dense colloidal suspension that is undergoing cyclic shear. Samples are deformed using a custom built interfacial stress rheometer that permits characterization of the sample microstructure while simultaneously measuring its bulk flow respnse (i.e. rheology). Contrary to intuition, we find that there is no chaotic progression in time (in other words, they do not evolve from elastic to reversibly plastic to irreversibly plastic). Instead, particles develop specific trajectories based on their position within the shear channel, and the strain amplitude. For example, above yield,  particles in the center of the channel are much more likely to have irreversible trajectories. Also, both plastically and elastically reversible particle trajectories transition to irreversible trajectories in later cycles (and vice versa); they sometimes change states. However, plastically and elastically reversible trajectories do not transition between each other. These observations are used to deduce the presence of a melting front, whose depth increases with strain amplitude. Based on a quantification of this depth we present an empirically determined strain amplitude scaling that quantifies plastic dissipation.

\section{Methods}

\begin{figure}[t]
\includegraphics{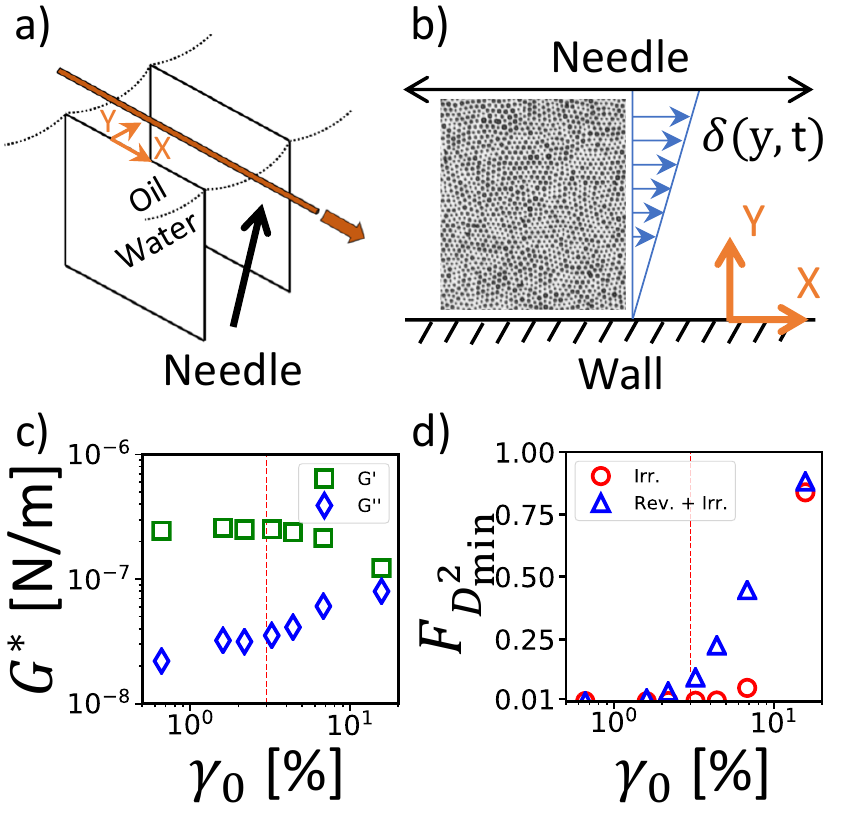}
\caption{\label{fig:1} Schematic of the system and background of data. a) Diagram of the interfacial stress rheometer including oil/water contact line pinned at the two glass walls and the axially displacing needle. b) A top view schematic with a description of coordinates and the idealized displacement field, $\delta(y,t)$. Also shown is an image of the particle micro-structure representing about 1/24th of a total image. The vertical edge is 250$\mu m$ long. Crystallized grain clusters may be observed, surrounded by expansive amorphous boundaries. c) Storage, G', and loss modulus, G'', as a function of strain amplitude $\gamma_0$, both showing inflection at the classic yield point of $\sim$3\% (\color{red}\textbf{- - -}\color{black}). d) Characterization of the fraction of particles displaying irreversible and reversible non-affine events. The total number of reversible and irreversible events diverge at the yield point.}
\end{figure}

We study the yield transition using an interfacial stress rheometer as shown in Fig.~\ref{fig:1}a. In short, a steel rod (230$\mu$m in diameter, 28.1mm in length), referred to as a needle, is placed at a water/decane interface. The interface is pinned on either side of the needle by glass walls (18mm long, 3.175mm spacing), ensuring the interface is planar. A monolayer of particles is also adsorbed at the interface (Fig.~\ref{fig:1}b). To shear the monolayer, the needle is driven sinusoidally by a uniform magnetic field, which is imposed by a pair of Helmholtz coils \cite{shahin86}. 

Rheological information is calculated by measuring the displacement of the needle using an inverted microscope and comparing to the imposed force. The effect of the interface on the needle is characterized by fitting fluid imparted drag and magnetic field imparted spring forces to the solution of a forced spring-mass-damper second order differential equation across measured needle displacement and proscribed force on the needle. The interface's effect is subtracted directly from the total observed response of a monolayer and interface, giving the storage and loss moduli, notated $G'$ and $G''$ respectively \cite{brooks99,reyn08}. 

To ensure rheological measurements are accurate, drag from the bulk fluid must be negligible compared to the drag from the interface. This ratio is calculated directly by the Boussinesq number, $Bq=|\eta^*|d/\eta_L$. $\eta^*$ is the observed complex viscosity, $d$ is the needle diameter, and $\eta_L$ is the liquid viscosity of the oil and the water, which is $\sim 10^3$ Pa s. Here the Boussinesq number is $\sim 10^2$ \cite{brooks99,reyn08}, so that in plane stresses are dominant.

Three systems of mono-layers, composed of non-Brownian particles are presented here. Table.~\ref{tab:table1} provides a summary of their differences. All three systems have crystaline grains with large amorphous swaths at the boundaries. However, the degree of crystalinity differs greatly between the three. For more information on disorder in these systems see \cite{keim13,keim14,keim15}. All are composed of mixtures of non-Brownian, sulfate latex spheres of nominal diameters 4.1$\mu$m and 5.6$\mu$m (Invitrogen). Sulfate charge groups coat the surface, creating an overall dipole-dipole repulsion force between particles \cite{park10}. These inter-particle forces are strong enough to create a stable material at the relatively low area fractions studied here, $\sim 31-43\%$. We refer to the monolayer as ``jammed'' in the sense that without shear, the individual particles do not undergo measurable changes in position --- let alone rearrangements. In addition to being jammed, this material is also soft, meaning that it can be deformed readily. An example image of monolayer A is shown in (Fig.~\ref{fig:1}b). Packings typically have small grains of a few particles with amorphous boundaries. Images span the space between a wall and the needle ($\sim 1000\mu m$) and include $\sim$40,000 particles. During each experiment imaging is carried out at 100-600 frames per cycle for up to 30 cycles. Features are identified and linked together using Trackpy \cite{allen18_tp}. The resultant trajectories are analyzed in several ways discussed below. The analysis presented in this paper is of the Bi-disperse monolayer with 50-50\% distribution, to serve as a demonstration. Final results, however, are shown for all three systems. Information about analysis of the other monoayers are available upon request.

\begin{table}[t]
\small
  \caption{\ A summary of the properties of the systems presented here, including dispersity, particle size ratios, sizes of particles, area fractions, $\Phi$. }
  \label{tab:table1}
  \begin{tabular*}{1.0\textwidth}{@{\extracolsep{\fill}}llll}
    \hline
    Dispersity & Ratio & Diameters & $\Phi$ \\
    \hline
     Bi-disperse & 50-50\% & 4.1, 5.6$\mu m$ & $\sim31$\% \\
     Mono-disperse & N/A & 5.6$\mu m$ & $\sim35\%$ \\
     Bi-disperse & 60-40\% & 4.1, 5.6$\mu m$ & $\sim43\%$ \\
    \hline
  \end{tabular*}
\end{table}

Here, frequency is held constant at $0.1Hz$ during all experiments with monolayer A. Strain amplitude, $\gamma_0$, is defined as needle displacement amplitude, $\delta_0$, divided by the distance between the wall and the needle. $\gamma_0$ is varied between 0.7\% and 17\%. This range fully traverses the yield transition, which is known generally to be near 3\% strain amplitude for many amorphous or glassy materials \cite{cubuk17}. Yielding is often designated based on an inversion of $G'$ and $G''$. As seen in Fig.~\ref{fig:1}c the inversion occurs near 3\% strain amplitude, consistent with previous findings. Further information about this system can be found in \cite{keim14}. 

As a touchstone to previous work reported in the literature $D_{2,min}$ calculations are presented. $D_{2,min}$ can be thought of as a quantification of a local deformation's non-linearity; i.e., it is the mean squared deviation of particle positions from a best-fit affine transformation over a time interval. We normalize this value by the square of the typical particle separation, $a$, and the number of neighbors considered (those within the two nearest neighbor shells, $~2.5a$). A non-affine event is characterized as a particle having a $D_{2,min}$ above $0.015$, a value used in simulations of amorphous solids. This threshold has been found previously to correspond to a disturbance in particle location of about $~0.1a$ \cite{falk98}.

\begin{figure}
\includegraphics{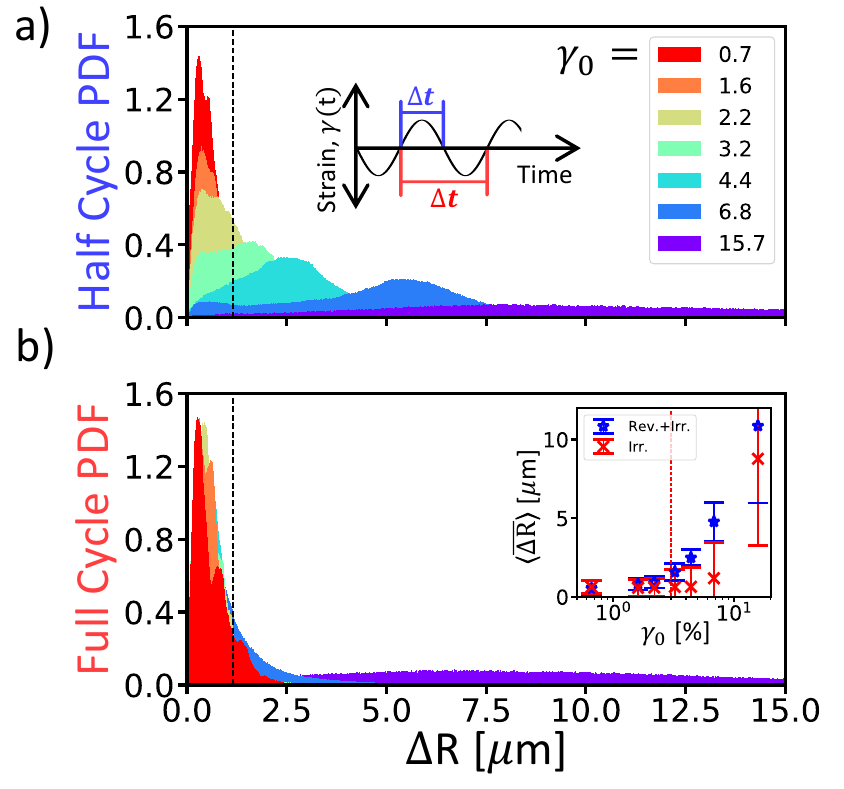}
\caption{\label{fig:2} Probability density function comparisons of particle displacement after a) half cycles and b) whole cycles over a range of strain amplitude. a) Particles fall within a small displacement below yield, slowly transition to a bi-modal distribution near yield, and finally nearly all particles escape above yield. A threshold (- - -) is included, found previously in simulation studies, $0.1a=~1.15\mu m$. Inset: visual representation of when half cycles and whole cycles are taken relative to strain, $\gamma_0$.  b) A separate transition is present, well above yield. Inset- the average displacement of particles that are to the right relative to strain amplitude over half and whole cycles. }
\end{figure}

In this paper, we measure $D_{2,min}$ over two time intervals: half cycles and whole cycles as shown in the inset of Fig.~\ref{fig:2}a. Whole cycle events are termed 'irreversibly plastic' because they indicate particles that have not returned to the positions they held at the beginning of the cycle. Half cycle events are also thought to detect irreversible plasticity, however they also detect a second type of event: 'reversible plasticity'. These events are characterized by particles that do not return to their original position after a half cycle, but do in fact return after an entire cycle. These particles typically trace limit-cycle trajectories as mentioned above. For further information on reversible and irreversible plasticity in this system and similar systems please see \cite{keim13,keim14}. Here we report the number of $D_{2,min}$ events averaged across steady-state cycles as a fraction of all particles observed, $F_{D_{2,min}}$ (Fig.~\ref{fig:1}d). 

To build a physical understanding of the types of non-affine events (reversible or irreversible) that are occurring, we measure several characteristics of each trajectory. One characteristic is the displacement of a particle over a half cycle and a whole cycle (see the inset of Fig.~\ref{fig:2}a for the time intervals used). From this information it is possible to determine weather any given particle has returned to its original position or not, using a threshold of $~0.1a$ as found in previous $D_{2,min}$ analysis \cite{falk98}. To be explicit, any particle that returns to within a tenth of the inter-particle spacing, $a$, has returned and a particle that does not return to within $0.1a$ has escaped. 

\begin{figure}
\includegraphics{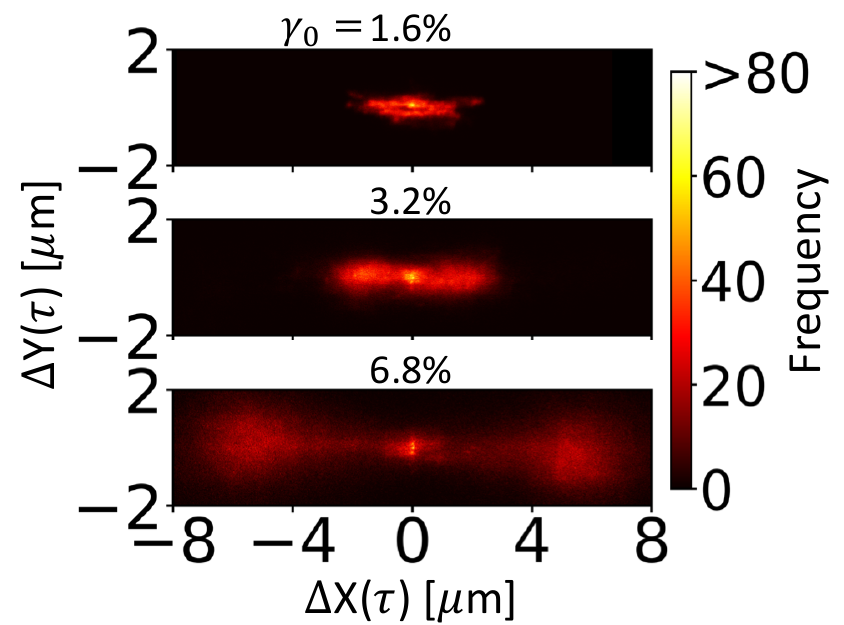}
\caption{\label{fig:3} Characterization of half cycle displacement vectors by x-y displacement-displacement Poincar\`{e} sections for three strain amplitudes. An additional attractor develops with increasing strain amplitude. Strain is increased from below yield (top), to near yield (middle), to above yield (bottom). Below yield, most particles return to their original position as expected from Fig.~\ref{fig:2}. Above yield, many particles do not return; their paths back to the origin are cut short, ending at periodic, chaotic points centered on the x-axis. These points grow outward with strain amplitudes above yield ($\sim$3\%). This can be seen in detail for all strain amplitudes in a video within the Supplemental Information. }
\end{figure}

\begin{figure}[!b]
\floatbox[{\capbeside\thisfloatsetup{capbesideposition={right,top},capbesidewidth=4cm}}]{figure}[\FBwidth]
{\caption{\label{fig:4} Above yield trajectories ($\gamma_0=$6.8\%). Trajectories are black with a red plus, (\color{red}+\color{black}), at the beginning of the cycle. For reference, local displacement is offset above in blue (\color{blue}---\color{black}). a-b) Trajectories dominated by mechanical noise. c) A low area example of a trajectory with arc length equal to the expected displacement ($L_N=0$). d) A high area example of a trajectory with $L_N=1.0$.}
}
{\includegraphics{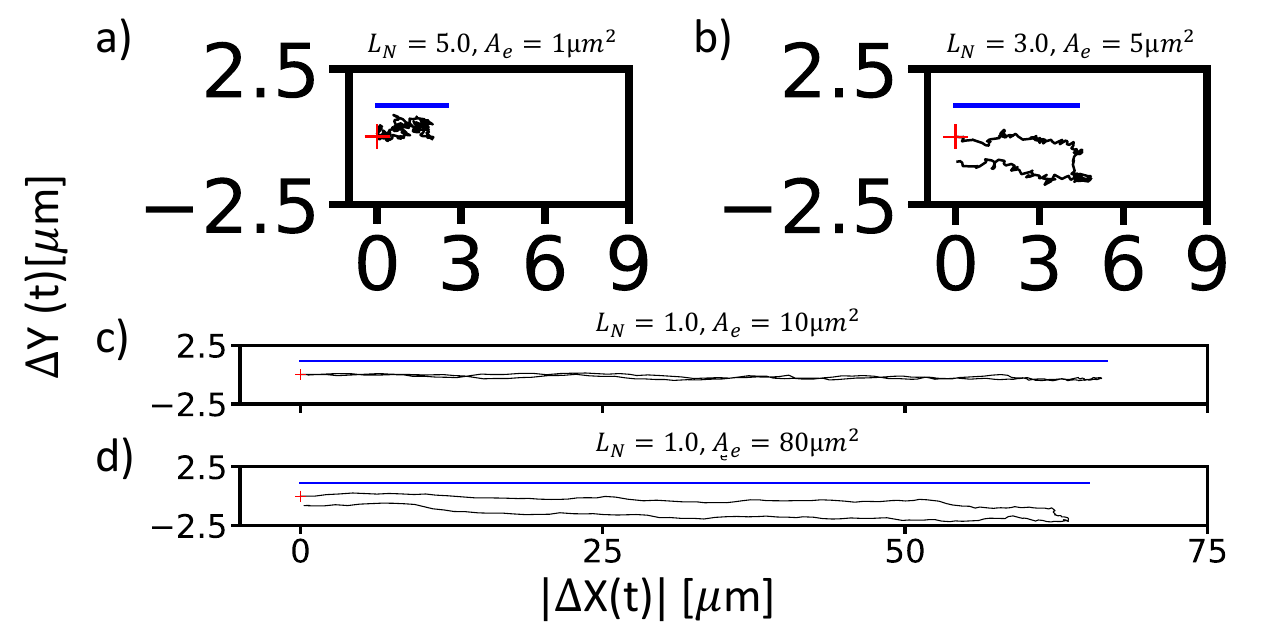}}
\end{figure}

\begin{figure}[b]
\includegraphics{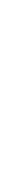}
\end{figure}

\begin{figure*}
\caption{\label{fig:5} Inverse normalized arc-lengths ($1/L_N$) and enclosed area (color bar) compared with the mean particle position between the needle and the wall. Irreversible particles are shown in black. a) Below yield the system is dominated by reversibly elastic, and irreversibly plastic particle trajectories. All trajectories have $1/L_N<1.0$, indicating that trajectories are long relative to the displacement field. This means they are dominated by mechanical noise. b) Near yield, plastically reversible particles emerge near the needle. Overall the $1/L_N$ shifts nearer to one (especially the plastically reversible particles) indicating a transition to low mechanical noise relative to affine displacements. c) Particles in the middle of the channel are exclusively plastically irreversible. Plastically reversible particles reach $1/L_N \sim 1.0$ indicating that these trajectories are completely dominated by background displacement, while simultaneously enclosing high area. It is worth noting that not a single particle is observed to have a $1/L_N >> 1.0$. }
{\includegraphics{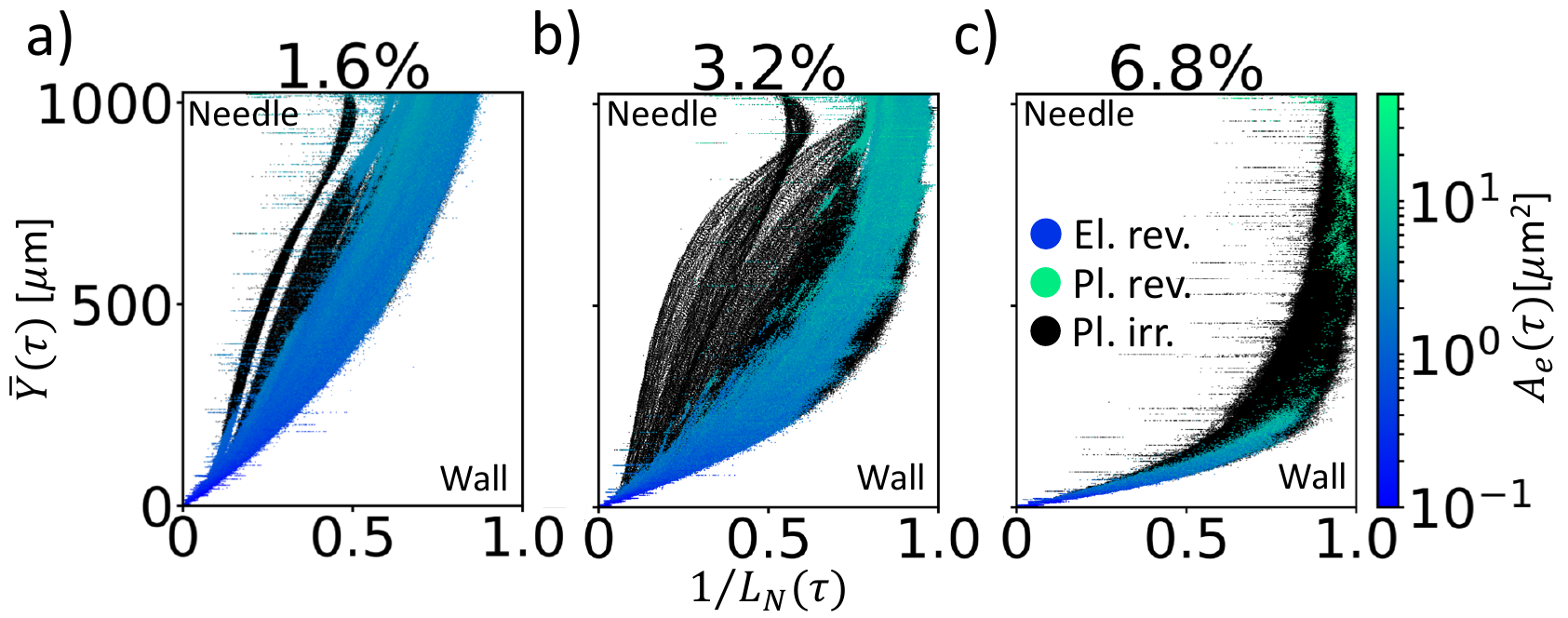}}
\end{figure*}

In addition, we calculate the area enclosed by trajectories of the particles that did not escape (the area of those that have escaped is defined as zero). This calculation is not as straight forward as it may initially seem; trajectories often intersect tens of times in a single half cycle. However, standard area calculation algorithms detect self intersections as a negative area and will not produce absolute area. Therefore, we have implemented a highly optimized algorithm that detects intersections and redefines sub-polygons that together make up the original area. Each sub-polygon's area is then calculated and summed. In addition to area, we calculate total arc-length.

Clusters are present within the enclosed area and arc-length phase space. To determine the relative numbers of particles transitioning from any given cluster to another, an algorithm must first be used to identify boundaries between clusters. The algorithm used in this paper is known as HDBSCAN (Hierarchical Density-Based Spatial Clustering of Applications with Noise) \cite{camp13,mcinnes17}. This clustering routine is strong at detecting clusters based on variations of density as well as distance within the chosen phase-space, and it is easily implemented in Python.

\section{Results}

We are interested in comparing the Lagrangian dynamics of each individual particle within the system; how these variables may change with position in the shearing channel, between successive cycles, and with strain amplitude. To gain this insight, we first investigate whether strain amplitude may be interpreted as a bifurcation variable affecting the propensity of particles to escape from their nearest neighbors. Fig.~\ref{fig:1}d, shows that the number of total non-affine events bifurcate at the yield point. To investigate further, Fig.~\ref{fig:2}a gives a normalized histogram of the displacement distance over half a cycle. Crucially, particles from the near yield case ($3.2\%$) show a bi-modal distributions about the noted threshold. Particles to the left have returned to their original positions, whereas those on the right have escaped their original particle positions. As strain amplitude is increased past yield, particles transition from below the threshold to above it. In the case of whole cycles (Fig.~\ref{fig:2}b), particles remain below the threshold, except when the system is well above yield. These trends (inset of Fig.~\ref{fig:2}b) are qualitatively similar to those of $D_{2,min}$ shown in Fig.~\ref{fig:1}d. Notably, half of the particles are seen to escape over half cycles at the yield point. Also, nearly all of the particles are seen to escape over whole cycles at the strain amplitude of equal elasticity and plasticity ($\sim 17\%$) seen in figure Fig.~\ref{fig:1}c.

A natural way to glean more information is to consider the components of the displacement vectors in each coordinate direction. To do this we plot Poincar\`{e} sections of spatial displacement (every half cycle) in Fig.~\ref{fig:3}. Plots of the remaining strain amplitudes are included in the supplemental information. Here we again find confirmation that there is a deviation of attractors above and below yield. It is seen that the attractor at the origin diminishes with strain amplitude, but is still present. The attractor representing escapes is visualized as a cloud of points and grows outward with the increase of strain amplitude. 

We have observed no evidence of structure within either attractor, therefore we believe these to be fully chaotic. Interestingly, the attractor that emerges at yield has two periodic points (see the supplemental material for supporting information about periodicity). Remarkably, the periodic points are directly centered on the $\Delta y=0$ axis. The whole cycle analysis shows two periodic points growing outward along the horizontal axis as well, with the caveat that they do not move outward as far as those shown in the half cycle cases (which is expected from Fig.~\ref{fig:2}b).

These results paint a picture of typical trajectories and their changes with strain amplitude. Particles predominantly move in the direction of needle displacement as expected. Moreover, particles that do not return are of a specific type: they are on track to return, but their trajectories get cut short by the end of the cycle. I.e. by the end of each half cycle of strain they have not returned \emph{yet}. Crucially, this implies that distance travelled by a particle is linked with the type of trajectory it creates (Fig.~\ref{fig:4}).

These findings inspire the inspection of a different phase space that can be thought of as a version of efficiency of dissipation. We recall that enclosed area is thought to correspond to energy dissipation. Therefore, it is natural to think of arc-length of an enclosed area as a way of measuring the efficiency of that energy dissipation. In other words, we measure how far a particle needs to travel to dissipate a certain amount of energy to its surroundings. However, particles will exhibit very different arc-lengths depending on how close they are to the wall; if a particle is very close to the wall it will hardly move at all. This leads us to normalize each trajectory's arc-length by the displacement that would be expected given its average y position, assuming a linear strain profile from the needle to the wall as shown in Fig.~\ref{fig:1}b. From geometry of similar triangles, this length is $2\gamma_0\delta(x,t)$. We define normalized arc-length as $L_N = \frac{L_{arc}}{2\delta_0 \langle Y \rangle}$.

\begin{figure}[!t]
\includegraphics{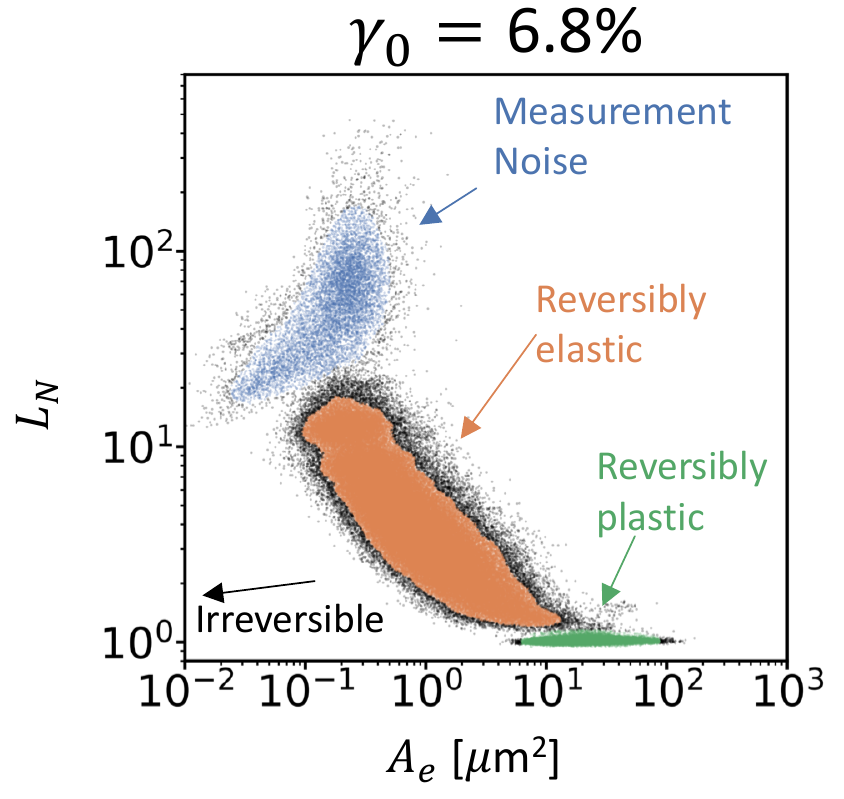}
\caption{\label{fig:6} A direct view of the efficiency space described. Normalized arc-length, $L_N$, is plotted against Enclosed area, $A_e$. Above yield clusters emerge that correspond to the reversibly plastic, in addition to the reversibly elastic cluster. Colored clouds of points demonstrate the HDBSCAN clustering algorithm employed for our data. Here, the strain amplitude is 6.8\%  }
\end{figure}

The efficiency spectrum introduced above is shown (indirectly) in Fig.~\ref{fig:5}. This set of plots show the mean particle position between the instrument needle and wall, $\overline{Y} (\tau)$, as a function of $1/L_N$, where $\tau$ is cycle number. We choose the inverse of $L_N$ for reasons apparent below. We show data for three strain amplitudes ranging from below yield (Fig. 5a, 1.6\%) to well above yield (Fig. 5c, 6.8\%). For the case below yield (Fig.~\ref{fig:5}a), particles trace out a wide range of $1/L_N$ relative to the expected displacement (based on needle displacement). However, each particle has an $1/L_N$ that is below one. This indicates that the arc lengths are long compared to the expected linear displacement field. These relatively long particle displacements are a result of erratic (i.e. non-smooth) particle paths. These erratic motions are due to small perturbations to the material's underlying energy landscape caused by small displacements of neighboring particles. This effect is commonly known as mechanical noise \cite{pine05}. An example of such a trajectory is shown in Fig. 4(b). Interestingly, the lowest values of $1/L_N$ are near the wall, implying that the effect of mechanical noise is much higher there than near the needle where displacements are largest. The enclosed area of trajectories is small relative to higher strain amplitude cases, which means that particles are predominantly elastically reversible in this case. 

Near yielding (Fig.~\ref{fig:5}b), we observe the appearance of trajectories that are reversible and plastic (green points), predominantly near the needle. $1/L_N$ shifts closer to one near the needle (and even the center of the channel), reflecting a decrease in the importance of mechanical noise relative to the low-strain case. This effect corresponds to the emergence of much higher enclosed areas, constituting plastic reversibility. Finally, well above yield (Fig.~\ref{fig:5}c), $1/L_N$ reaches one at the needle meaning that mechanical noise is nearly negligible (for the case of particle motion) and trajectories are mostly smooth (see Fig. 4c,d). Additionally, these reversibly plastic trajectories reach enclosed areas that are almost an order of magnitude larger than in lower strain amplitudes. Crucially, the particles in the center of the channel become nearly completely irreversibly plastic. 

These results seem to imply that yielding is characterized by particles that dissipate energy with a minimized arc-length. In contrast, particles below yield dissipate very little energy while exhibiting large arc-lengths (high mechanical noise; see trajectories in Fig.~\ref{fig:4}a-b). Randomized particle motions due to mechanical noise dominate the particle system below yield. As strain is increased, this motion becomes smaller relative to the overall strain-driven displacement. Once above yield, the effect of mechanical noise is negligible relative to the local displacement, resulting in arc-lengths that are smaller relative to the local affine displacement field. Crucially, arc-lengths smaller than the linear displacement are not observed (Fig.~\ref{fig:4}c). Once this limit is reached, enclosed areas grow and the system begins to dissipate energy (Fig.~\ref{fig:4}d).

\begin{figure}[!t]
\includegraphics{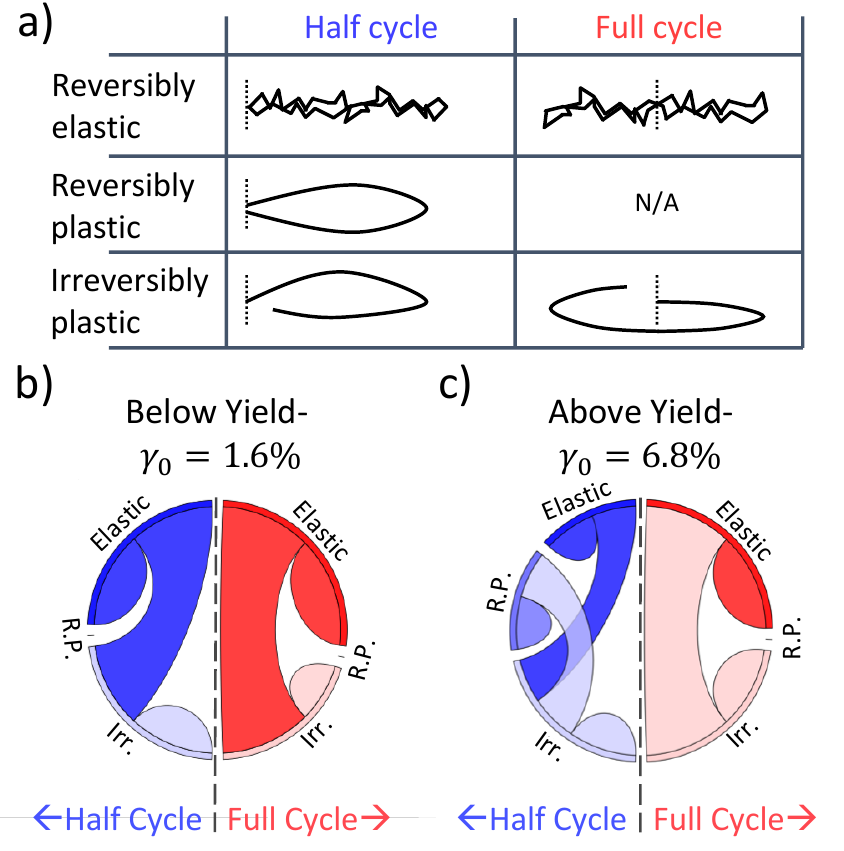}
\caption{\label{fig:7} a) Cartoon representation of different particle trajectories corresponding to those shown in Fig.~\ref{fig:4}. b\&c) Chord diagram representations of particle's inter-cycle transitions between apparent clusters within Fig.~\ref{fig:5}a\&c. Widths of cords at either end represent the log of the numbers of particles transitioning from that state. Color of each cord corresponds to the state that has more particles transitioning. b) Below yield for both the half cycles and whole cycles there is no presence of the reversible plastic cluster. c) Above yield, half cycles exhibit a reversibly plastic cluster, whereas the whole cycles do not. The reversibly and irreversibly plastic states do not exchange particles. }
\end{figure}

A direct view of the efficiency space described above is shown in Fig.~\ref{fig:6}, where $L_N$ is plotted against $A_e$. Here separate clusters are immediately apparent. Above yield, a large $A_e$ cluster emerges, which corresponds to the reversibly plastic state described above. Because irreversible particles do not enclose an area, they are off of the logarithmic horizontal axis. 

\begin{figure}[t]
\includegraphics{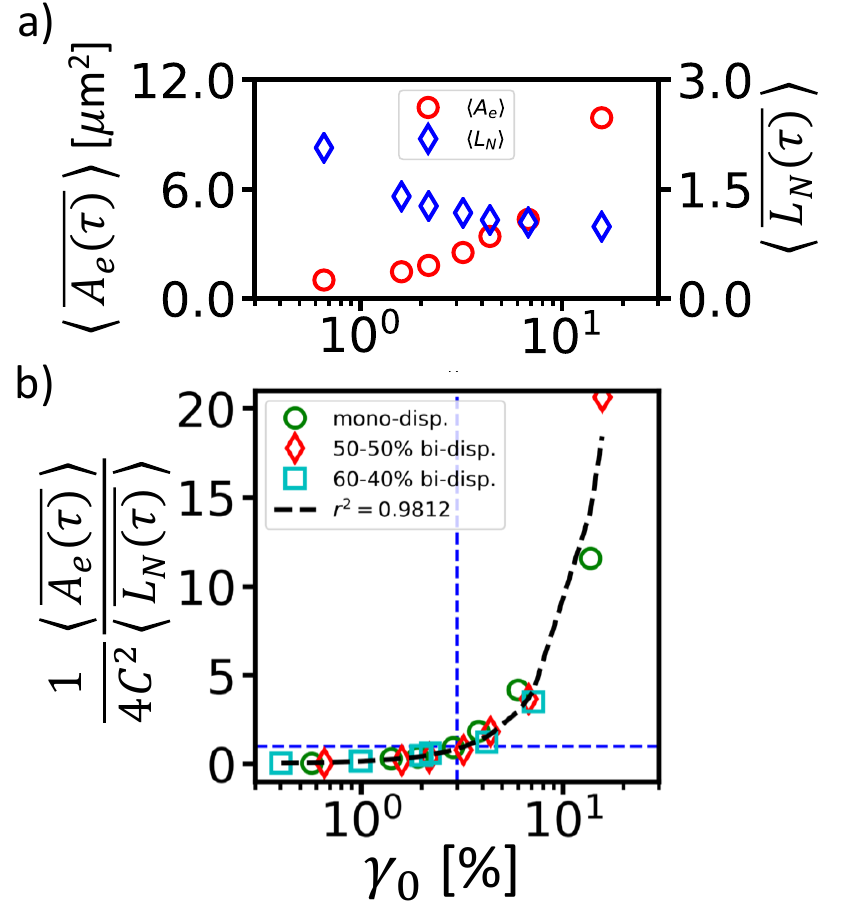}
\caption{\label{fig:8} Trends in average enclosed area and normalized arc-length as a function of strain amplitude. a) With increasing strain amplitude, average enclosed area, $\langle A_e \rangle$ grows rapidly. With strain amplitude, average normalized arc-length, $\langle L_N \rangle$, drops monotonically toward an asymptote at unity. b) Taking the square root of the ratio, $\langle A_e \rangle / \langle L_N \rangle$ we find a linear collapse between three colloidal systems of various amounts of disorder. Moreover, this collapse passes through unity at the yield point ($\gamma_0 \sim 0.3 $).}
\end{figure}

The presence of attractors and states brings up the question of how particles may be transitioning from one half cycle to the next. We perform a cluster analysis to answer this question. The efficiency spectrum of enclosed area and normalized arc-length provides a convenient way to determine clusters of elastically reversible, plastically reversible, and irreversibly plastic trajectories. Once these clusters are determined, questions of how many particles transition between states from one cycle to the next can be answered quantitatively. 

The colored clouds of points shown in Fig.~\ref{fig:6} reflect the detection of clusters by the HDBSCAN algorithm. One observation is that the algorithm does not designate quite all of the particles around the borders of the designated clusters; it leaves out 0.30\% of the total points. This is due to large differences in the density of points within the designated clusters and the outer fringes; there are narrow, 'loose and fuzzy' edges. A second note is that there is a cluster labeled as 'measurement noise'. These points correspond to particles within a few particle diameters of the wall. Trajectories are highly noisy for these particles because there is optical disturbance from the wall. This cluster comprises 0.32\% of the overall points. These points are discarded. 

To display our results quantitatively, we have elected to use two chord diagrams (Fig.~\ref{fig:7}b,c). These plots give a quick but quantitative assessment of the numbers of transitions from one state to the next. Thickness of chords indicate the logarithm (base 10) of the number of particles transitioning out of one state into another; e.g. a chord that is wide on one end and narrow on the other indicates more particles leaving the wide state than from the narrow state. This functions almost like an arrow, where the widths indicate the logarithm of the number transitioning. 

As a summary of each state found, we have included a table of cartoon depictions of the observed particle paths (  Fig.~\ref{fig:7}a) We note here that even though reversibly plastic trajectories are not observed over whole cycles, the trends indicate that they may occur at higher strain amplitudes than are presented here. The first observation to draw from Figure~\ref{fig:7} is that particles readily transition between the elastic and plastically reversible states both over half and whole cycles. A second observation is that both elastically reversible and plastically reversible particle trajectories do not transition between each other, implying that they are separate populations. This is a break from intuition, which would have that particles transition from elastically reversible to reversibly plastic and then to plastically irreversible. 

Two findings have been made: reversibly plastic and elastic states exist independently of each other in time, and the particles fall into these states based on where they are within a melting front between the needle and the wall. The depth of this melting front, as characterized by the normalized arc-length, increases with strain amplitude. Crucially, we have seen that the enclosed area increases rapidly above the yield point and that the normalized arc-length decreases monotonically towards unity. In Fig.~\ref{fig:8}a we present both of these quantities averaged in time and space, relative to strain amplitude. Remarkably, taking the ratio of these two quantities reveals a parabolic relationship with strain amplitude, as shown in Fig.~\ref{fig:8}b. This functionality suggests a relationship of the type
\begin{equation}
\frac{\gamma_0}{\gamma_{cr}} = \frac{1}{2C} \left( \frac{\langle \overline{ A_e (\tau)} \rangle}{\langle \overline{ L_N (\tau)} \rangle} \right)^{0.5}
\label{eq_scale}
\end{equation} 
where $\gamma_{cr} = 3.0\%$ is the critical strain amplitude indicating yield. The two is included to account for the fact that there is material being sheared on either side of the needle. Eq.~\ref{eq_scale} is a dimensionless scaling, quantifying plastic loss as a function of strain amplitude. That is, this equation uses Lagrangian particle dynamics to describe the yield transition. 

Enclosed area of limit cycles is related to dissipated energy, which has been measured previously to increase rapidly beyond the yield point \cite{keim13,keim14}. Therefore these results lead us to conclude that the ratio, $\langle \overline{ A_e (\tau)} \rangle $/$ \langle \overline{ L_N (\tau)} \rangle$ (having units of $\mu m^2$) is a direct Lagrangian measure of dissipation within the entire system. ($L_N$ appearing here in the denominator is the reason we above plot against $1/L_N$ in Fig.~\ref{fig:5})

To explain the origin of the coefficient $C$ we must consider the subtlety of plasticity. As particles rearrange, they must squeeze past each other. This relaxation process results in local forces acting between particles. These forces on the bulk scale give rise to fluctuating normal forces on the needle and walls. This is known as Reynolds dilatancy \cite{lu18}. The walls are fixed in space, whereas the needle is constrained only by contact with the particles on either side of it. These normal force fluctuations cause the needle to displace at low frequencies as material on both sides of the needle relax. However, this length-scale must adhere to a value that depends on the particle interaction strength and the sizes of particles themselves. In these experiments, we observe this value to be $C = 1.04 \pm 0.15 \mu m [95\%]$.

\section{Discussion and Conclusions}

In this paper we investigated the Lagrangian dynamics of constitutive particles within a 2D soft jammed material spanning strain amplitudes from below yield to above. Our first finding is that the average displacement distance of particles from the beginning of a cycle to the end increases monotonically with strain amplitude in a qualitatively similar way to the number of non-affine events (both reversible and irreversible)(Fig.~\ref{fig:1}d and Fig.~\ref{fig:2}b inset). Moreover, these displacements are predominantly in the direction of shear, which is linked to particles falling short of returning to their original positions (Fig.~\ref{fig:3}). This in turn implies particle motions have slowed, dissipating energy; thus there is importance to measuring area and arc-length as proxies for energy dissipation and a material's ability to dissipate energy (efficiency) respectively. 

As has been noted previously, \cite{pine05} particles below yield are found to display erratic motions about a mean path (reversibly elastic). The mean path is the expected local displacement due to a linear strain field. These erratic motions are caused by mechanical noise from perturbations of the energy landscape (Fig.~\ref{fig:4}a\&b and Fig.~\ref{fig:5}). However, above yield particles near the shearing surface are found to instead have smooth trajectories with limit cycle trajectories (reversibly plastic)(Fig.~\ref{fig:4}c\&d and Fig.~\ref{fig:5}). Moreover, any particle that does not maintain these limit cycles, will not return to its initial position (irreversibly plastic). These non-returning particles are found everywhere throughout the channel below and near yield. However, well above yield, they are the only variety of trajectory found in the center of the channel. 

More broadly, these may be signs of what is happening within the energy landscape; as particles transition from rough to smooth trajectories at the yield point, the energy landscape is transitioning from a fixed state with small perturbations, to being actively changed, but in ways that reverse as strain inverts. Irreversible particle trajectories represent local, permanent changes to the landscape and become dominant well beyond yield. The magnitudes of these changes has been shown in our system (Fig.~\ref{fig:2}b inset) and in simulations \cite{kawa16} to suddenly transition in strain amplitude, similar to a first order state transition. Strain is a temperature-like variable. Other indications that the yield transition is generally of the first order have been observed recently in simulations and theory \cite{ozawa18} and experiments \cite{denis15, naga14}. 

\begin{figure}[t]
\includegraphics{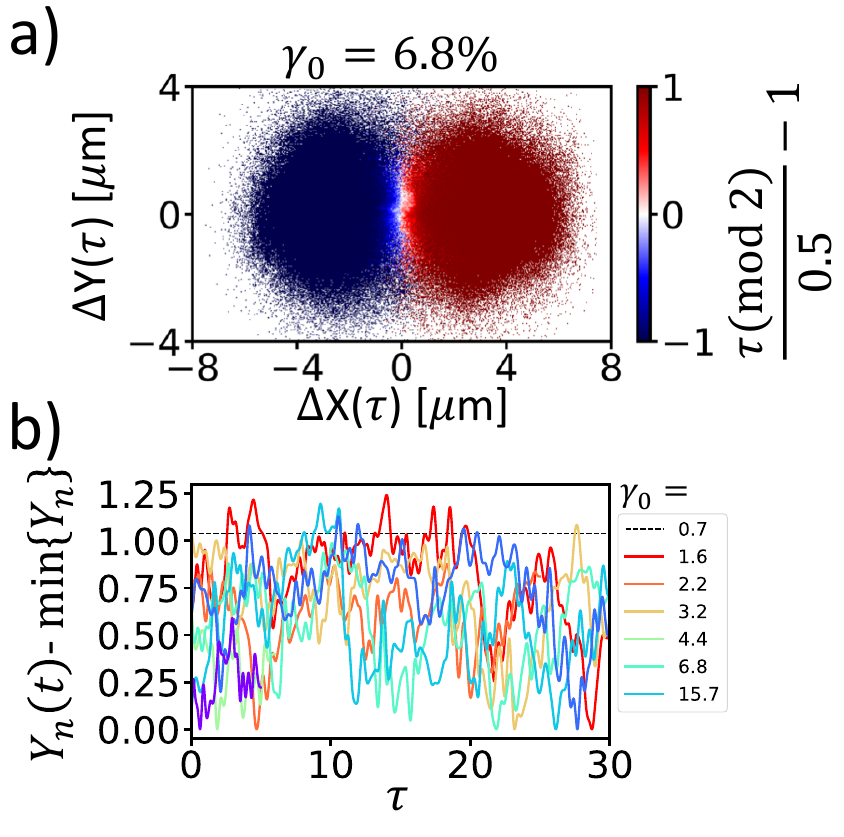}
\caption{\label{fig:9} a) Horizontal and vertical axis are a Poincar\`{e} section well above yield. Color represents a binned average of the remainder of half cycle count divided by two, shifted so that even numbers of half cycles are blue and odd half cycles are red. The left half of the attractor is composed of even half cycles and the right is made up of odd half cycles. Therefore particles must be oscillating between both points. b) The range in the the displacement of the needle in the direction normal to shear plotted against cycle, $\tau$, for all strain amplitudes. The average difference between the highest and lowest displacement is about one here. $C = 1.04 \pm 0.15 \mu m [95\%]$ is shown as a dashed gray line (-).}
\end{figure}

We find that a quantification of this yield surface is the ratio of the average enclosed area (increasing with strain amplitude) and the average normalized arc length, $L_N$, (falling with strain amplitude). Area serves as a quantification of the energy dissipated by a limit cycle \cite{Strog94,schreck13,regev13}. We posit that the inverse of $L_N$ is a measure of how efficiently a particle is dissipating energy relative to a given strain amplitude. Using both of these concepts, we present a non-dimensional quantification of plastic dissipation based entirely on the Lagrangian dynamics of the constituent particles (Eq.~\ref{eq_scale} and Fig.~\ref{fig:8}). 

More generally, this quantification works because it captures variations in particle response between the stationary and shear surfaces. While enclosed area of trajectories is not an applicable concept outside of oscillatory systems, $1/L_N$ is applicable and should be measured in other systems. In this paper, we describe a spatial transition in $1/L_N$ that is associated with the yield transition, and the location of this transition penetrates farther from the shearing interface the more strain is increased. This is reminiscent of a melting front, and consistent with dynamics observed in dry and immersed granular systems \cite{koma01,kats10,hous15,hous16}. In particular, steadily-sheared granular systems exhibit: a decreasing shear rate away from the shearing interface; a fluidized layer at the shearing interface where particles move ballistically; and a transition associated with a critically-low shear rate to caged dynamics characteristic of glassy materials \cite{ferd18, hous16}. The thickness of the fluidized layer has been found to be proportional to the applied shear rate \cite{hous15}, and the melting front has been identified as a yield surface marking the transition to (athermal) granular creep. Seeking similarities in the melting-front dynamics of these systems and our experiments is an exciting next step, which will help to reveal whether the creep-flow transition in granular systems is a similar state transition to what we observe here.   


We have presented echoes within a colloidal system under oscillatory stress of phenomena observed in granular systems under steady shear. This introduces evidence that may help answer a tantalizing question: how is the particulate behavior of amorphous systems of greatly variable length scales, various interaction forces and complicated shapes related to each other? And how is that related to the bulk response? Some observations have been made previously: yield strain within amorphous materials as a whole is $\sim3\%$ \cite{cubuk17}; in granular systems, dimensionless strain rates are related directly to the volume fraction \cite{boyer11, Jerol19}. The results presented here indicate that particle displacement relative to the expected strain may be a useful tool in understanding the response of amorphous materials more generally. This parameter is accessible and should be investigated further in a myriad of systems.

\section{Appendix}

Within the main body of the text, two attractors were shown in Fig.~\ref{fig:3} and discussed. The point at the center of all Poincar\`{e} sections is an attractor corresponding to particle returning to their original positions. Two additional points develop with increasing strain amplitude, constituting a bifurcation. In the main body it was stated that these points are periodic within the same attractor. In Fig.~\ref{fig:9}a we supply evidence. By numbering sequentially, $T$, each half cycle of every particle's trajectory we display where each particle is during even and odd numbered half cycles. In the Poincar\`{e} section in Fig.~\ref{fig:9}a we show that a binned average of the remainder of T corresponds to each point. Particles on the right overwhelmingly are of odd half cycles, whereas particles on the left are overwhelmingly even half cycles. This means that particles are switching from the right to the left points periodically throughout a cycle. Therefore, the attractor is made up of two periodic points. 

Within the main body of the text, it is mentioned that the characteristic displacement length, C, of the shearing tool perpendicular to shear is $\sim 1.0 \mu m$. In Fig.~\ref{fig:9}b we plot the y-position of the needle minus the minimum position versus time (in units of cycles). The highest peak for any given strain amplitude corresponds to an estimate of C. The average of these peak heights across all three samples is $C = 1.04 \pm 0.15 \mu m$ over a 95\% confidence interval.

\section*{Acknowledgements}

We thank Troy Shinbrot and S\`{e}bastien Kosgodagan Acharige for fruitful discussions. This work was This work was supported by ARO W911-NF-16-1-0290 and the Penn NSF MRSEC (DMR-1120901). 

\bibliography{references}
\bibliographystyle{rsc}

\end{document}